\documentclass{article}

\usepackage{arxiv}
\pdfoutput=1 

\usepackage[utf8]{inputenc} 
\usepackage[T1]{fontenc}    
\usepackage{hyperref}       
\usepackage{url}            
\usepackage{booktabs}       
\usepackage{amsfonts}       
\usepackage{nicefrac}       
\usepackage{microtype}      
\usepackage{lipsum}		
\usepackage{graphicx}
\usepackage[round]{natbib}
\usepackage{doi}
\usepackage{amsmath}

\title{A Method for Estimating Individual Socioeconomic Status of Twitter Users}

\author{\hspace{1mm}\small{Yuanmo He} \\
	\small{Department of Methodology}\\
	\small{London School of Economics and Political Science}\\
	\small{London, WC2A 2AE, United Kingdom} \\
	\small{\texttt{\href{mailto:y.he54@lse.ac.uk}{y.he54@lse.ac.uk}}} \\
	\And
	\hspace{1mm}\small{Milena Tsvetkova} \\
	\small{Department of Methodology}\\
	\small{London School of Economics and Political Science}\\
	\small{London, WC2A 2AE, United Kingdom} \\
	\small{\texttt{\href{mailto:m.tsvetkova@lse.ac.uk}{m.tsvetkova@lse.ac.uk}}} \\
}

\renewcommand{\shorttitle}{A Method for Estimating Individual Socioeconomic Status of Twitter Users}

\hypersetup{
pdftitle={A Method for Estimating Individual Socioeconomic Status of Twitter Users},
pdfsubject={Computers and Society},
pdfauthor={Yuanmo He, Milena Tsvetkova},
pdfkeywords={socioeconomic status, correpondence analysis, Twitter, measurement, social media, cultural capital},
}

\begin{document}
\maketitle

\begin{abstract}
    The rise of social media has opened countless opportunities to explore social science questions with new data and methods. However, research on socioeconomic inequality remains constrained by limited individual-level socioeconomic status (SES) measures in digital trace data. Following Bourdieu, we argue that the commercial and entertainment accounts Twitter users follow reflect their economic and cultural capital. Adapting a political science method for inferring political ideology, we use correspondence analysis to estimate the SES of 3,482,652 Twitter users who follow the accounts of 339 brands in the United States. We validate our estimates with data from the Facebook Marketing API, self-reported job titles on users’ Twitter profiles, and a small survey sample. The results show reasonable correlations with the standard proxies for SES, alongside much weaker or non-significant correlations with other demographic variables. The proposed method opens new opportunities for innovative social research on inequality on Twitter and similar online platforms.
\end{abstract}


\section{Introduction}
Socioeconomic status (SES), a concept that describes people’s social and economic position relative to others, is one of the most fundamental concepts in social science, underlying major areas of research such as health, education, psychology, sociology, and public policy \citep{diemer_best_2013, krieger_measuring_1997, oakes_measurement_2017, rodriguez-hernandez_socio-economic_2020}. Some researchers focus on measures of SES, in an attempt to capture the social stratification of modern society \citep{chan_class_2007,hauser_socioeconomic_1997,savage_new_2013}, while others investigate how SES relates to other life outcomes and thus propagates socioeconomic inequality. We know, for example, that people’s SES affects their physical and mental health \citep{adler_socioeconomic_1994,dohrenwend_socioeconomic_1992}, political participation \citep{brady_beyond_1995, milligan_does_2004}, the size and diversity of their social circle \citep{campbell_social_1986, marsden_core_1987}, and their access and use of information and communication technologies \citep{van_deursen_digital_2014, van_deursen_third-level_2015, hargittai_digital_2008}. Most notably, people’s SES is highly predictive of their children’s SES, outlining the major pathway through which inequality is transmitted, social mobility constrained, and advantage accumulated across generations \citep{diprete_cumulative_2006, sirin_socioeconomic_2005}.

Most of the existing quantitative research on SES and socioeconomic inequality relies on statistical models of survey, census, and administrative record data. The recent rise of computational social science (CSS), however, offers opportunities to study socioeconomic inequality with an entirely different set of tools and data---applying text analysis, network analysis, or machine learning methods to web, mobile, or satellite “digital trace” data \citep{lazer_computational_2009}. For example, CSS researchers have combined night-time maps with high-resolution daytime satellite images to estimate poverty in regions with poor administrative data \citep{abitbol_interpretable_2020, jean_combining_2016}. Scientists have also analysed aggregate data on Google searches and daily usage patterns of Twitter to predict unemployment claims before official statistics are released \citep{choi_predicting_2012, llorente_social_2015}. Others have used social media and mobile network data to link economic development to social capital, showing that individuals who live in areas with a high local development index tend to have more diverse networks \citep{eagle_network_2010} with bridges that span greater geographic distances \citep{norbutas_network_2018}.

These CSS studies on socioeconomic inequality, however, are conducted at the level of geographic units. Large-scale individual-level analyses using digital trace data are less common since researchers rarely have access to users’ demographic and financial information. One notable exception is a unique dataset that couples mobile phone communication with bank transaction history for a subsample of the population of a Latin American country \citep{leo_socioeconomic_2016, leo_correlations_2018, luo_inferring_2017}. Another prominent exception is a recent research collaboration between high-profile social scientists and Facebook, granting access to rich individual information for millions of the online social network’s US users \citep{chetty_social_2022}. Data like these, however, tend to be proprietary and not easily accessible. 

To address this gap, computer scientists have developed various methods for inferring demographic attributes from openly available digital-trace data; however, very few of these concern SES, social class, and their indicators: income, education, and occupation \citep{hinds_what_2018}. Researchers are yet to find an effective, theoretically grounded, and scalable method to infer the individual-level SES of online users. Such a method will allow linking measures of SES to the detailed records of everyday decisions, behaviors, opinions, and interactions that digital-trace data offer. The resulting research will provide population-level natural-setting observations of the daily reproduction of socioeconomic inequality. A better understanding of how limited financial resources and education may drive self-defeating behaviour, strain interactions with others, or restrict access to valuable information will empower us to tackle inequality from the bottom up, complementing top-down legislative and policy reforms.

The current paper addresses the identified gap in the literature by outlining a method to estimate the SES of individual Twitter users. Twitter is a social media platform with 1.3 billion accounts and 330 million monthly active users, where 500 million tweets are posted per day \citep{brandwatch_60_2020}. It is one of the most popular social media platforms used for CSS research: the number of Twitter-related studies is consistently growing \citep[see reviews by][]{karami_twitter_2020, yu_bibliometric_2020, mccormick_using_2017}. The public messaging aspect of Twitter provides valuable opportunities for researchers to observe behaviours, social interactions, and networks with a minimum obtrusion, in real-time, at a low cost, and on a large scale. Moreover, Twitter offers a well-developed application programming interface (API) that makes the data more accessible compared to other popular digital platforms (e.g., Facebook, Instagram, TikTok). 

Nevertheless, it is hard to infer Twitter users’ socioeconomic status. Twitter does not have a designated field that requires socioeconomic information. Some Twitter users state their occupations in their profile description field, but few disclose this information accurately or at all \citep{sloan_who_2015}. Reviews on the topic show that existing studies on estimating the SES of individual Twitter users are scarce and disparate, and most of them have methodological limitations \citep{ghazouani_assessing_2019, hinds_what_2018}.

In this paper, we present a method to estimate the SES of individual Twitter users from the commercial and entertainment accounts they follow on the platform. The method parallels an established political science approach that uses correspondence analysis to estimate Twitter users’ political ideology from the politicians and news media they follow \citep{barbera_birds_2015, barbera_tweeting_2015}. In accordance with Bourdieu’s \citeyearpar{bourdieu_distinction_1984} multidimensional definition of social class, the proposed measure of SES aims to capture a combination of economic and cultural capital. As economic and cultural practices may differ in different countries, we here present the method using popular brands in the US and US Twitter users only. With the information from the Twitter accounts of 339 brands and their followers, we are able to estimate the SES of 3,482,652 users. We validate our estimation with brand consumer statistics from Facebook, self-described occupation from thousands of Twitter profiles, and survey responses on education and income from a small sample of Twitter users. Although further fine-tuning and external validation will be desirable, our preliminary results indicate that the method promises to become a valid and useful measure of SES for Twitter users.

\subsection{Measuring SES: from survey data to Twitter}
The idea to approach modern societies as strata or segments of SES groups is one of the most fundamental and deeply rooted ideas in sociology, tracing its origins back to Durkheim, Marx and Weber. Yet, 150 years later, the problem of how to define and measure SES is still contested and unresolved. There are debates regarding whether SES is unidimensional or multidimensional and what to include in the measure \citep{chan_understanding_2019, chan_class_2007, flemmen_class_2019, hauser_socioeconomic_1997, savage_new_2013}. Nevertheless, in practice, SES is often viewed as a “shorthand expression” for variables indicating certain aspects of SES such as income, education, and occupation \citep{hauser_socioeconomic_1997}. These variables typically appear among standard demographic variables included in surveys, making it convenient to link SES to various other measures used in social science. SES is thus often measured or represented by one or a combination of these variables. The popular approaches to measure SES include using a univariate proxy such as just income or just education, a composite measure which incorporates income, education, and occupation such as Duncan’s Socioeconomic Index \citep{duncan_socioeconomic_1961} and the Nam-Powers occupational status scores \citep{nam_variations_1965}, or an occupation-based class schema such as the Erikson-Goldthorpe-Portocarero (EGP) class schema \citep{erikson_constant_1992}. 

Therefore, the most obvious approach to infer Twitter users’ SES would be to estimate their income, education, or occupation. For instance, researchers can automatically extract job titles from users’ profile description, rely on some sort of human validation to exclude inaccurately labelled jobs, and then link the titles to income or occupational class \citep{ghazouani_assessing_2019, sloan_who_2015}. One can also obtain occupation from the LinkedIn links users include in their profile or tweets \citep{abitbol_optimal_2019}. The problem is that very few users state their job title or include a link to their professional accounts in their profile descriptions. Thus, the approach severely reduces the size of the sample to tens of thousands at most and potentially biases it towards individuals who act in official capacity, such as journalists, promoters, or politicians. 

Using another data mining approach, researchers can estimate income or wealth by linking geo-located accounts and tweets to average house value or income at the census block level \citep{abitbol_optimal_2019, park_economic_2018}. Similarly, however, users who disclose their geo-location are rare \citep{jiang_understanding_2019}. Around 30-40\% of Tweets contain some profile location information, but the profile location tends to be at the region, state, city, or county level; the more granular geo-tagged tweets only make up one to two percent \citep{twitter_get_2020}.

Employing more sophisticated machine learning techniques, other studies estimate SES with supervised methods trained on various Twitter features \citep{ghazouani_assessing_2019}. However, stemming from computer science, these studies do not engage sufficiently with social theory to justify the features and outcome variables used in the models \citep[e.g.,][]{filho_inferring_2014, moseley_user-annotated_2014, preotiuc-pietro_analysis_2015, volkova_predicting_2015, volkova_mining_2016}. For example, in one of the most cited papers on estimating Twitter users’ SES, \cite{preotiuc-pietro_studying_2015} employ the Bayesian non-parametric framework of Gaussian Processes to predict user income and occupational class from a large bag of features, including the number of followers, proportion of retweeted tweets, and the average number of tweets per day, among others, together with psycho-demographics, emotions, and word topics inferred from textual analysis of the user’s posts. The authors train their model on the income and occupational class associated with the job titles retrieved from user descriptions. However, because they use too much information in estimating the SES with complex models, there is limited usage for the estimates. The paper also relies on aggregate-level information (income associated with job titles) to estimate individual SES without individual-level validation; this is another prevalent problem in the existing literature \citep[e.g.,][]{aletras_predicting_2018, ardehaly_mining_2017, filho_inferring_2014}.

We contribute to existing efforts to estimate individual SES on Twitter by proposing an alternative unsupervised learning method. Political scientists have successfully used this method to estimate Twitter users’ political ideology \citep{barbera_birds_2015, barbera_tweeting_2015} and here, we adapt it to estimate SES. The method relies on correspondence analysis, a simple dimensionality-reduction technique that is already familiar to cultural and Bourdieusian sociologists, and is thus more accessible to less methodologically savvy social scientists than alternative complex supervised machine learning approaches such as Bayesian Gaussian Processes \citep{preotiuc-pietro_studying_2015} or neural graph embeddings \citep{aletras_predicting_2018}. The method uses minimal, commonly available, and easily accessible information about Twitter users’ followings and employs fast off-the-shelf estimation algorithms, making it data economical, computationally efficient, and scalable. Specifically, the method yields SES estimates for millions of users compared to prior studies’ benchmarks in the neighbourhood of 50,000 \citep{aletras_predicting_2018, sloan_who_2015}. Finally, as we argue in the next section, the method relies on assumptions that are firmly embedded in classical sociological theory: Bourdieu’s \citeyearpar{bourdieu_distinction_1984} habitus theory. This renders the method relevant and useful for various strands of sociological research; it also directly responds to the recent call for better integration of data, measurement, and theory in computational social science \citep{wagner_measuring_2021, lazer_meaningful_2021}. Parenthetically, the proposed method aligns with the latest budding approaches to studying SES and culture with graph embeddings \citep{kozlowski_geometry_2019, taylor_concept_2020}, as recent research shows the mathematical and interpretive similarity between correspondence analysis and embedding methods \citep{van_dam_correspondence_2021}.

\subsection{Measuring SES as economic and cultural capital with cultural interests and consumer preferences}
\citet{bourdieu_distinction_1984} viewed an individual’s SES as a function of their economic, cultural, and social capital. Economic capital refers to material resources such as wealth and income, cultural capital refers to the valued competence of engaging with cultural goods, and social capital refers to the network of contacts and connections that could be useful when needed. People’s social position and the capital they possess shape how they act in and perceive the social world. Bourdieu calls this sense of orientation towards the social world habitus. The habitus manifests itself in people’s everyday social practices and becomes concretely visible in people’s cultural tastes and preferences. This manifestation may not be necessarily conscious and intentional but is nevertheless strategic, in the sense that it serves to distinguish one’s social status and to distance oneself from other groups \citep{bourdieu_distinction_1984}. Thus, on the one hand, people’s upbring, education, and social surroundings shape their taste and cultural interests to be coherent within their own SES group. On the other hand, the exclusive nature of taste, which rejects cultural interests that are inconsistent with one’s own SES, divides people into distinct and divergent SES groups. 

Bourdieu mainly focused on the role of cultural tastes and cultural consumption for social distinction. \citep{veblen_theory_2017} made a similar argument about distinction but instead emphasised the role of economic purchases. Using the concept of conspicuous consumption, Veblen argued that people tend to use material goods and leisure activities to demonstrate their SES to others. In other words, distinction could materialize not only via cultural tastes but also in preferences for consumer products and brands.

Naturally, Bourdieu’s theory has been challenged, qualified, and extended since then. Most notably, while Bourdieu identified an accentuated taste stratification and classification in France, others have shown that, in the United States for example, individuals of higher social status tend to be “cultural omnivores,” espousing broader and more eclectic cultural tastes \citep{holt_does_1998, peterson_understanding_1992}. Similarly, the recent notion of inconspicuous consumption suggests that people with more wealth and cultural capital actually tend to be more subtle and less ostentatious consumers \citep{berger_subtle_2010, eckhardt_rise_2015}. Thus, more recent research challenges the idea that low versus high SES can be neatly mapped onto low- versus high-brow cultural tastes and basic versus luxury consumption. Nonetheless, it leaves intact two main assumptions that are crucial for our argument here: 1) people express their SES via their cultural interests and consumer preferences, and 2) people in similar SES tend to have similar cultural interests and consumption preferences.

Consequently, we argue that we can use the cultural interests and consumer preferences people declare on social media to estimate their SES. Specifically, we assume that Twitter users manifest their economic and cultural interests with the accounts they follow on Twitter. Many commercial and entertainment brands, including retailers (supermarkets, department stores, apparel), chain restaurants, news sources, sports associations, and TV shows, have official Twitter accounts. The brands use these accounts to share news, promote products and events, and interact and engage with fans, and users who value this information are more likely to follow these accounts. Marketing research shows that 35\% of Twitter users in the US use Twitter to follow brands \citep{werliin_new_2020}. Academic research shows that the main motivations for Twitter users to follow brands are incentives (discounts, coupons, promotions, etc.), information (to know more about products), social interactions (to interact with brand representatives or like-minded people), and attitudes toward brands \citep{kwon_brand_2014}. These motivations align well with the framework of the habitus: preferences, interests, interactions, and attitudes represent different aspects of a person’s orientation toward the social world, which reflects their socioeconomic background. Following consumer brands (e.g., retailers and chain restaurants) represents a combination of economic and cultural preferences: the price tag of the good or service reflects the economic constraints a person faces, and the associated quality and style represent the person’s cultural taste and lifestyle. Following media and entertainment brands (e.g., news sources, sports associations, and TV shows) mainly represents cultural interests. Even if we don’t know which brands represent higher economic and cultural capital, we can cluster users who tend to follow similar brands and project them onto a line, which will serve as our SES scale. This is the basic idea behind the method we propose below.

As a matter of fact, Bourdieu himself used a similar idea and a related method to demonstrate his concept of multidimensional social space. In his influential book Distinction \citep{bourdieu_distinction_1984}, Bourdieu applied a dimensionality-reduction technique known as multiple correspondence analysis (MCA) on a survey sample of the French population in the 1960s containing data on income, occupation, and engagement in various cultural activities (e.g. reading, going to concerts, visiting museums). The technique allowed him to position individuals, occupations, and cultural activities on a two-dimensional graph. Bourdieu argued that the first dimension represents the overall volume of economic and cultural capital and the second dimension represents the contrast between economic and cultural capital \citep{bourdieu_distinction_1984, weininger_pierre_2005}. Despite ongoing debates on the measurement of cultural capital and the relation between cultural interests and SES \citep{peterson_changing_1996, prieur_emerging_2013, reeves_how_2019}, a recent study reaffirmed the utility of using Bourdieu’s method to establish social space and measure SES as a combination of economic and cultural capital \citep{flemmen_social_2018}. 

In contrast to Bourdieu’s surveys, we rely on the economic and cultural interests people reveal on social media.  Computer scientists, political scientists, and psychologists have already used these data to extract various information about online users: demographic characteristics, political ideology, and psychological traits, as well as other private and sensitive information \citep{hinds_what_2018}. For instance, the researchers behind the myPersonality study apply supervised learning methods on participants’ “likes” for Facebook groups to show that sexual orientation, ethnicity, religious and political views, personality traits, intelligence, happiness, use of addictive substances, parental separation, age, and gender can be predicted with relatively high levels of accuracy \citep{bachrach_your_2014, bi_inferring_2013, kosinski_private_2013, youyou_computer-based_2015}. More relevantly for us, political scientists utilize an unsupervised learning method to infer users’ position on the left-right ideological spectrum based on the Twitter accounts of politicians, political parties, media outlets, and journalists the users follow \citep{barbera_birds_2015, barbera_tweeting_2015} or the Facebook pages of politicians they like \citep{bond_quantifying_2015}. The method uses correspondence analysis (which is related to Bourdieu’s MCA) on the users and the official accounts they follow to project their position on a continuous linear scale. Below, we outline how the method can be adapted to estimate user SES. 

\section{Method}
The method relies on two sets of social media users: the accounts, public pages, or fan groups of consumer brands and cultural products and the individuals who follow, subscribe, or otherwise positively engage with them. It uses correspondence analysis (CA) to map the associations between the brands and users onto a two-dimensional space and then estimate the SES of the brand/user from its coordinates in the first dimension. Based on our theoretical framing, we assume that the prime reason for a user to follow a brand is SES proximity, in the sense of congruent economic preferences, cultural interests, and lifestyle. Therefore, the first dimension from CA that explains the most variance of the user-brand matrix is a valid representation of the users and brands’ SES. The use of CA is identical to political science approaches for estimating political ideology from Twitter followings and Facebook page likes \citep{barbera_tweeting_2015, bond_quantifying_2015} and in principle similar to the Multiple Correspondence Analysis (MCA) conducted by Bourdieu himself \citep{bourdieu_distinction_1984, flemmen_social_2018}. 

CA is a multivariate method to summarise and visualise the associations between rows and columns of a two-way contingency table as the positions between points in a low-dimensional space \citep{greenacre_correspondence_2017}. The low-dimensional space is identified so that the variance of the original matrix is explained by the dimensions in descending order. Since the first two dimensions explain most of the variance, the output of CA is often a two-dimensional plot. In our case, we use the first dimension to obtain measures on a continuous SES scale but the method could be adapted to use the first two dimensions and assign SES according to a discrete class-based schema.

For $\mathbf{N}$ representing a binary matrix with $I$ users as rows following $J$ brands as columns, CA is conducted through the following main steps (Greenacre 2017).  

First, we compute the matrix $\mathbf{S}$ of standardised residuals:
\[\mathbf{S = D_r(P - rc)D_c}\]
where $\mathbf{P} = \frac{1}{\sum_{i=1}^{I}\sum_{j=1}^{J}N_{ij}}\mathbf{N}$ is the binary data matrix transformed into proportions, $\mathbf{r}$ and $\mathbf{c}$ are the row and column weights with $r_i = \sum_{j=1}^{J}P_{ij}$ and $c_i = \sum_{i=1}^{I}P_{ij}$, and $\mathbf{D_r}$ = diag($1/\sqrt{ \mathbf{r} }$) and $\mathbf{D_c}$ = diag($1/\sqrt{ \mathbf{c} }$) are the diagonal matrices with diagonal entries equal to the inverses of the square roots of the weights. This step ensures the model captures the associations between rows and columns in a way that does not depend on row or column sums. In essence, it accounts for the fact that some users are more active and some brands are more popular in general.

Second, we calculate the singular value decomposition of $\mathbf{S}$:
\[\mathbf{S = UD_{\alpha}V^T}\]
where $\mathbf{U}$ and $\mathbf{V^T}$ are the left and right singular vectors of $\mathbf{S}$, which are orthogonal and hence $\mathbf{UU^T = V^TV = I}$, and $\mathbf{D_\alpha}$ is the diagonal matrix of singular values in descending order $(\alpha_1 \ge \alpha_2 \ge \cdots)$. In other words, we now represent the information in $\mathbf{S}$ with two coordinate matrices ($\mathbf{U}$ and $\mathbf{V^T}$) and a scaling matrix ($\mathbf{D_\alpha}$). Put simply, this step finds the low-dimensional space that best fits the original matrix in terms of least-squares approximation.

Finally, we project all rows and columns onto the plane by computing the standard coordinates: $\mathbf{G_r = D_rU}$ for rows and $\mathbf{G_c = D_cV}$ for columns. As the original data matrix $\mathbf{N}$ lists users in rows and brands in columns, the row coordinates $\mathbf{G_r}$ in the first dimension give the estimated SES of the users, and the column coordinates $\mathbf{G_c}$ in the first dimension---the estimated SES of the brands. Lastly, we standardize the estimates to have a normal distribution with a mean of 0 and a standard deviation of 1, which aids the interpretation of the estimation. Since CA captures the relative positions of the users and brands, the interpretation of the estimated SES should focus on the values relative to other values in the whole sample rather than the absolute values. For example, an estimated user SES of –1 means that the user has an SES that is one standard deviation lower than the average user SES in the sample.

We note that CA also allows projecting data points (users or brands) not used in the original estimation onto the same subspace. To do this for a new brand, for example, we take the standardized column with the users that follow it $\mathbf{n'} = \frac{\mathbf{n}}{\sum_{i=1}^{I}n_i}$ and compute $\mathbf{g} = \mathbf{n'}^T\mathbf{G_r}$. Similarly, we can map new users (Barbera et al. 2015).

\section{Data}
To test the validity of this method, we use the official Twitter accounts of a group of consumer brands and the followers of these accounts. Data collection and research for the study were approved by the University Ethical Review Board and the complete list of brands and their Twitter accounts required to replicate the results is available in the Supplementary Table 1. 

To identify the brands, we first selected six domains that cover various forms of daily material and cultural consumption: supermarkets and department stores, clothing and speciality retailers, chain restaurants, newspapers and news channels, sports, and TV shows. We then used Wikipedia lists, YouGov popularity rating lists, and media reports on TV shows’ audience \citep{maglio_tv_2016, wikipedia_list_2020, yougov_most_2018} to identify the most prominent brands in the US. From these, we selected the ones that have a Twitter account with more than 10,000 followers. We only included accounts with a large number of followers to ensure the accounts can contribute to the analysis. Further, for international brands, we included only their US accounts, whenever available. We thus started with 341 brands.  

Using the Twitter Search API \citep{twitter_get_2020} and the wrapper function in R developed by \citet{barbera_pablobarberaecho_chambers_2020}, we then obtained the full list of followers for these 341 brands till May 2020, yielding 191,790,786 users who follow at least one of the brands. To guarantee that we have sufficient information to characterize a user, we excluded users who follow fewer than five brands, which resulted in 23,567,268 users. Next, we used the users’ profile data to further delete inactive users and potential bots. We kept users who had sent at least 100 tweets, have at least 25 followers, and had sent at least one tweet in the first five months of 2020. This selection left 4,436,095 users. 

Finally, we were able to exclude some users who are not in the US based on the “location” field of their Twitter profile. We opted to exclude, rather than include users based on location data because these data are inconsistent and rarely available. For users who provide their location, some are easily identified just using text selection, as they put in a country or state name. For those who only put a street or city location, we used the Google Geolocation API \citep{google_developers_overview_2020} to match the street or city with the country. After excluding users whose location is not in the US, there are 3,482,657 remaining users. After pruning the users, two brands (“Red Mango” and “Saatva Mattress”) were left with only 0 and 1 followers, while the other brands had at least 1000. Since these two brands would not be informative for the analysis, we deleted them and then selected the users who follow at least five brands in the new sample. In the end, the sample contains a matrix of 339 brands and 3,482,652 users.

To improve the validity of the estimates, we conduct the analysis in two steps. First, we use CA on a maximally informative subset to identify the low-dimensional space and then, we project all users and brands to the space to estimate everyone’s SES. Specifically, for the first step, we select “informative users” who follow at least one brand from each of the six domains (supermarkets \& department stores, clothing \& speciality retailers, chain restaurants, newspapers \& news channels, sports and TV shows), resulting in 158,441 users. Then we select the “informative brands” followed by at least 1000 of the “informative users,” resulting in a 158,441 $\times$ 303 matrix (in comparison, the full matrix is 3,482,652 $\times$ 339). 

We conduct CA on this subset using the \emph{ca} package in R \citep{nenadic_correspondence_2007}. After confirming that the results are interpretable with a simple qualitative check, we use them to first project the coordinates for the rest of the brands, and then project the coordinates for the rest of the users. We use code from Barbera \citep{barbera_pablobarberaecho_chambers_2020, barbera_tweeting_2015} to do the projections. 

\section{Results and Validation}
\begin{figure}[ht]
	\centering
	\includegraphics{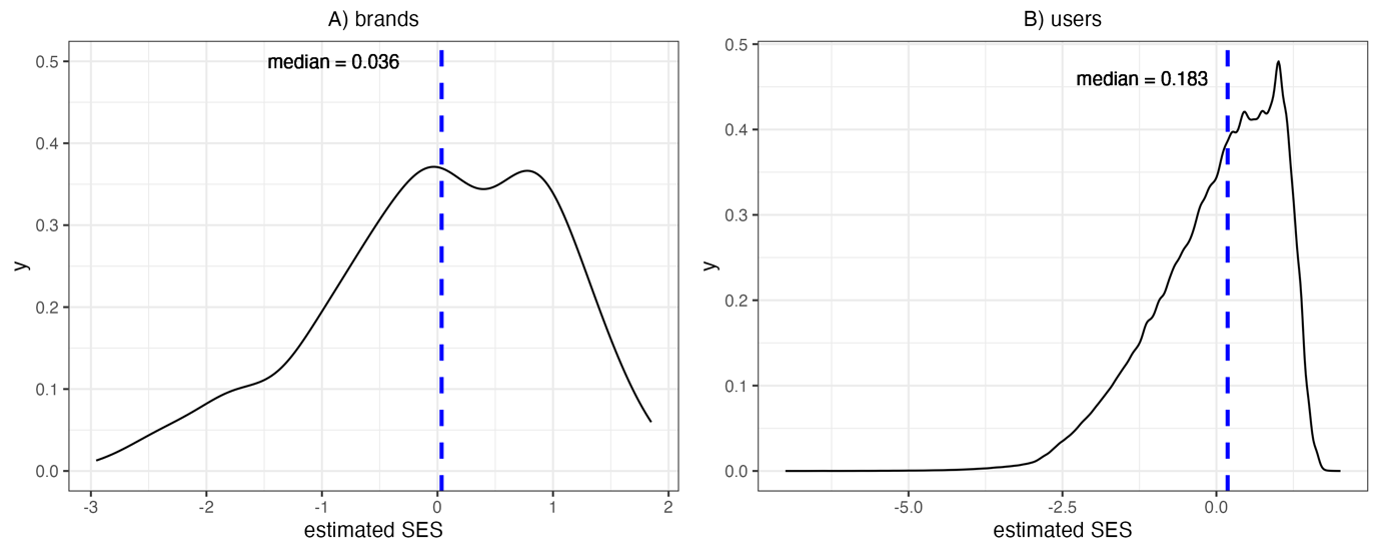}
	\caption{Density plot of the estimated SES for A) 339 brands and B) 3,482,657 users who follow them on Twitter.}
	\label{fig:fig1}
\end{figure}

Figure \ref{fig:fig1} depicts the density distributions of the estimated SES for the brands in our sample and the users who follow them on Twitter. The estimated SES for the brands ranges from $-2.95$ (\emph{hushpuppies\_usa}) to $1.85$ (\emph{soulcycle}), with a median of $0.036$. For the users, the estimated SES ranges from $-7.00$ to $2.02$, with a median of $0.183$. It is evident that both distributions are skewed towards middle-to-high SES. The skew for individuals corresponds well with the results from the nationally representative survey by Pew Research Centre showing that Twitter users are more educated and have higher income than the general US population \citep{wojcik_how_2019}. 

To validate the estimates, we bring in data from various sources and conduct analyses at both the aggregate and individual levels. Our first step is to establish convergent validity. First, we confirm the qualitative interpretation of the brands’ SES and compare our estimates with aggregate statistics on the educational level of the brands’ marketing audience obtained from Facebook. Second, we quantify the extent to which, on aggregate, the SES estimates correlate with the mean salary and occupational class for a subsample of users who include an occupational title in their Twitter profile information. Third, we estimate the extent to which the individual SES estimates predict education and income in a small survey sample of Twitter users. Our next step is to confirm divergent validity, namely, that the SES estimates are not measuring other related demographic variables. We use again data from Facebook, Twitter users, and the survey sample to confirm that the SES estimates are to a much lesser extent associated with age, gender, race, political ideology, and urban/rural residence.

We note that since SES is a composite concept, and our measure is operationalized to capture this multi-facetedness, we do not expect a perfect correlation between our SES estimates and any single one of the simple measures of education, occupational class, or income. Yet, neither can we rely on another composite measure such as the SEI as a ground-truth benchmark to measure our success against: as we mentioned in the introduction, the sociological community has not coalesced to a universal understanding of SES. Our primary aim here is to prove the existence of a meaningful signal in the proposed measure and stimulate further research that could better isolate, filter, and amplify this signal. 

\subsection{Validation of brand SES}
\begin{figure}[ht]
	\centering
	\includegraphics{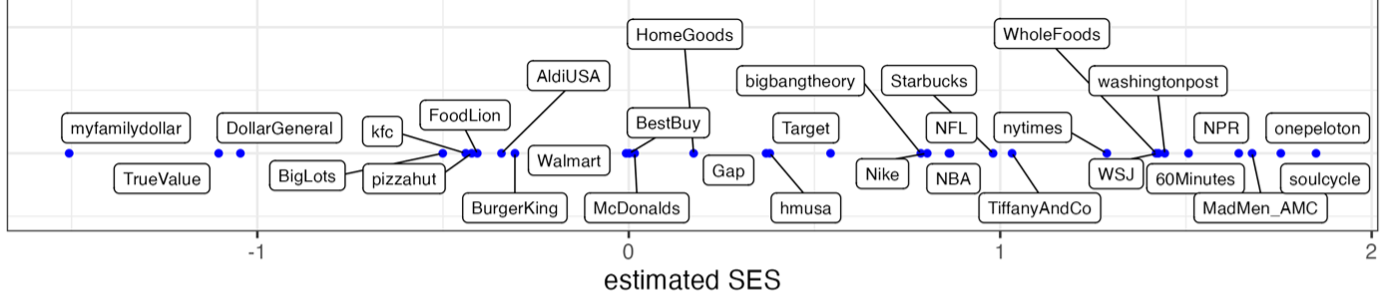}
	\caption{Estimated SES for a selected group of popular brands.}
	\label{fig:fig2}
\end{figure}

We begin by qualitatively sense-checking the SES estimates for brands. Figure \ref{fig:fig2} shows the estimates for a selected group of popular brands, while Supplementary Table 2 lists all estimates. The lower end of the scale has discount store chains such as \textit{Family Dollar}, \textit{Dollar General}, and \textit{True Value}. Slightly higher, there are fast food restaurant chains such as \textit{Pizza Hut}, \textit{KFC}, and \textit{Burger King}, and inexpensive stores and supermarket chains such as \textit{Big Lots} and \textit{Aldi}. The next band, constituting the first hump of the bimodal distribution visible in Figure 1A, contains many essential and/or large businesses: \textit{McDonald’s}, \textit{Walmart}, \textit{Best Buy}, \textit{Home Depot}, \textit{Old Navy}, \textit{Toys “R” Us}, etc. Then, there are average priced supermarket and clothing brands such as \textit{Target}, \textit{H\&M}, and \textit{Gap}. The most populated SES band (the second peak in Figure 1A) has the brands that one could argue are universally popular, such as \textit{Nike} for clothing, \textit{NFL} and \textit{NBA} for sports, the \textit{Big Bang Theory} for TV shows and \textit{Starbucks} for coffee chains. The higher end has iconic middle to elite class brands such as \textit{Whole Foods}, chic and expensive exercise brands \textit{Peloton} and \textit{Soul Cycle}, and the TV show \textit{Mad Men}, which in 2010 was reported to have 48\% of its viewers with household income of more than \$100,000 \citep{szalai_cable_2010}. The higher end also includes national newspapers such as \textit{The New York Times}, \textit{The Wall Street Journal}, and \textit{The Washington Post}. This result corresponds well with \citep{chan_class_2007}, which shows that national newspapers tend to be read by people with higher social status. 

In a next step, we validate the brands’ estimated SES quantitatively with data from the Facebook Marketing API \citep{facebook_marketing_2021}. Prior research on migration, health, urban crime, and digital inequalities \citep[e.g.,][]{araujo_using_2017, fatehkia_using_2018} demonstrates that the Facebook Marketing API can be an effective tool for obtaining population-level demographic estimates. With tailored targeting criteria, the API provides the number of users an ad can reach per month on Facebook. We use the targeting criteria to choose an interest, for example, \emph{soulcycle}, and find the number of active users whose highest earned degree is high school diploma, Bachelor’s degree, and Master’s or higher and who express interest in \emph{soulcycle} in the US, from which we then calculate the proportion of \emph{soulcycle}’s audience with different educational levels. We recognize that the audience on Facebook and Twitter is not entirely the same; expressing interest in a brand on Facebook and following a brand on Twitter may also represent different motives. Nonetheless, the Facebook audience data provide valuable insights into the brands’ audience composition and thus offer a useful reference for the validation of our measurement.

There are multiple educational levels in the Facebook data, including categories such as “in university” and “some degree”. For clarity, we only choose three levels that represent the full completion of a degree. Seven brands (\emph{FinishLine, GNCLiveWell, GreysABC, Gap, LEVIS, MakitaTools, CodeBlackCBS}) in our sample have an audience size of 1000 universally in all educational levels, which may mean Facebook does not have a reasonable estimate of the audience size for these brands. Further, no suitable data are available for four brands (\emph{moen, Hanes, thehill, WestworldHBO}). Therefore, we exclude these brands for this part of the analysis, resulting in 328 brands. Figure 3 plots the proportion of the brand’s Facebook audience at the specified educational level against the brand’s estimated SES according to our method. A small number of the brands’ Twitter screen names are shown alongside their points and to aid visibility, these are chosen for plot areas with low density of observations. Panel A) shows a negative association between the brand’s estimated SES and the proportion of users in the brand’s Facebook audience whose highest earned degree is a high school diploma (Spearman’s $\rho = -0.464$, $p < 0.001$), while panel C) shows a positive association between the estimated SES and the proportion who hold a Master’s or higher degree ($\rho = 0.444$, $p < 0.001$). Panel B) shows a somewhat lower but still positive association between the estimated brand SES and the proportion of users among the brand’s audience whose highest degree is Bachelor’s ($\rho = 0.320$, $p < 0.001$).

\begin{figure}[ht]
	\centering
	\includegraphics{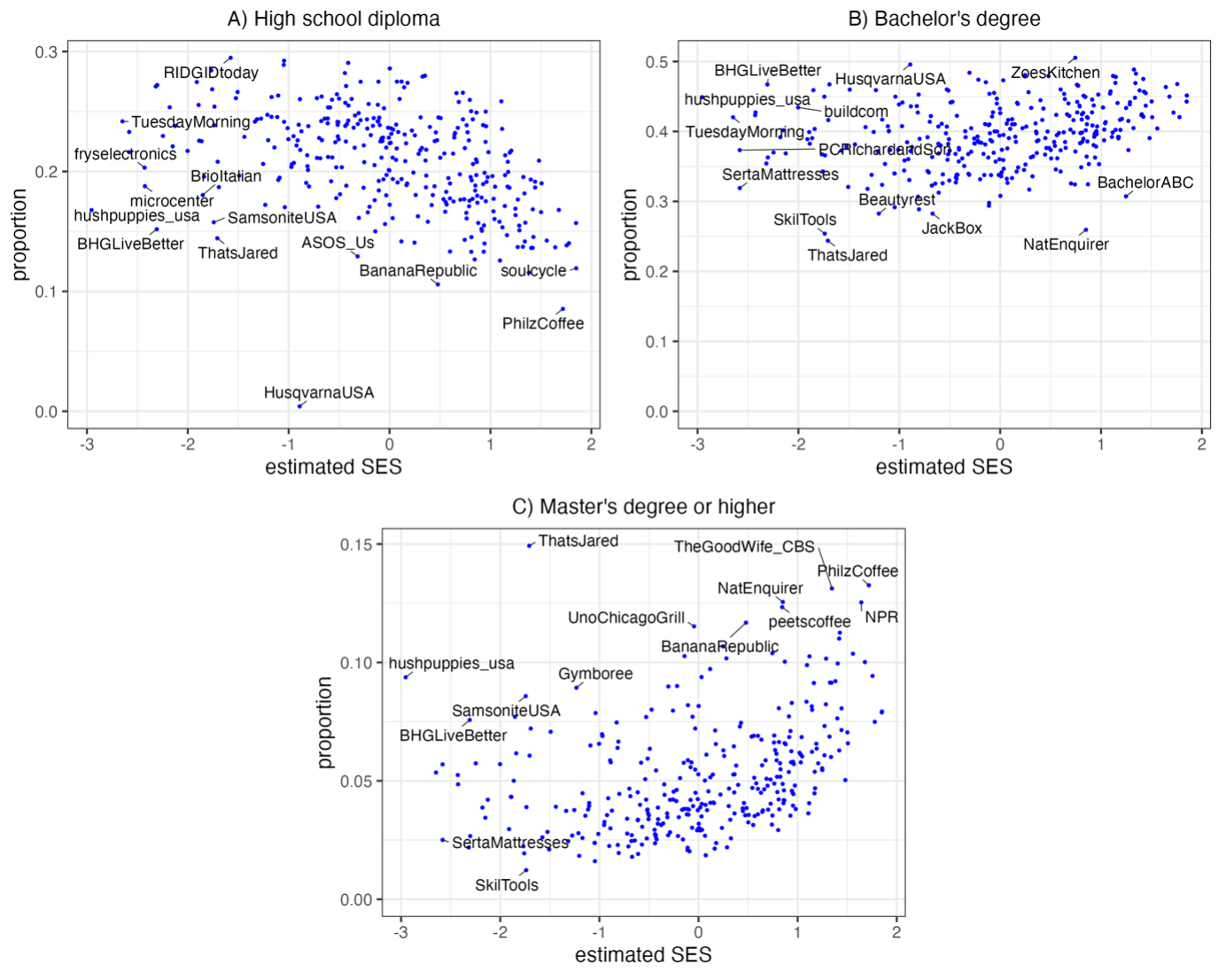}
	\caption{Relation between the educational composition of the brands’ Facebook audience and the brands’ estimated SES.}
	\label{fig:fig3}
\end{figure}

The patterns in the plots and the correlation statistics show that the brands with higher estimated SES tend to have significantly smaller audience of at-most high school graduates, significantly larger audience with Master’s or higher degrees, and somewhat larger audience of Bachelor degree holders. The latter represent the largest and most diverse audience on Facebook, so it is expected that their expressed interests in the brands are not as informative as the other education levels. These trends together suggest that the audience of the brands with higher estimated SES have higher average educational level than the audience of the brands with lower estimated SES. In sum, the proposed method positions consumer and media brands along an SES scale in ways that resonate with common knowledge and convincingly capture the educational level of the brand’s social media audience.

\subsection{Validation of user SES with self-reported job titles}
We next assess whether the SES estimates for users are valid too, starting at the aggregate level. We do this by identifying a set of common and informative job titles mentioned in Twitter profiles and comparing the income and occupational class associated with the job title to the average SES estimates for the Twitter users who state this job title in their profile description. Essentially, we quantify how the estimates by our SES measurement method compare on average to those by another prominent approach that relies on self-disclosed job titles  \citep{sloan_who_2015}. 

We complete the following steps to identify and match job titles. We first find job titles from different occupational social classes from the UK’s Standard Occupational Classification \citep{ons_standard_2020} and note their class. We choose the UK’s SOC instead of the US’s SOC because it has more specific job titles and is closer to the well-established Goldthorpe Class Scheme \citep{erikson_constant_1992, rose_national_2005}. Then we use text selection to search for the job titles in the profile descriptions of all users in our Twitter sample. We only include the job titles that return more than 50 users. To minimise the number of wrongly labelled titles, we include an additional filter: we manually inspect ten randomly sampled descriptions for each job title to identify text structures that contribute to mislabelling and then filter out the titles that match the text structures identified. After this filtering, we also delete two titles (tailor and waitress) that have fewer than ten cases. In the 2020's version of the UK SOC scheme, there are nine occupational social class levels, where a lower number means a higher occupational social class \citep{ons_standard_2020}. We try to include job titles from all nine classes, but job titles in some classes are harder to match with profile descriptions than others. After the text selection, we search the job titles in the “May 2019 National Occupational Employment and Wage Estimates” table on the website of the \citet{us_bureau_of_labour_statistics_may_2020}. We only include job titles that make sense in the US context and note their mean annual salaries. The outlined procedure resulted in a sample of 42,099 users matched with 50 titles, which we use as our validation set. Supplementary Table 3 lists the selected titles and their mean annual salary in US dollars, grouped by their occupational social class. 

\begin{figure}[ht]
	\centering
	\includegraphics{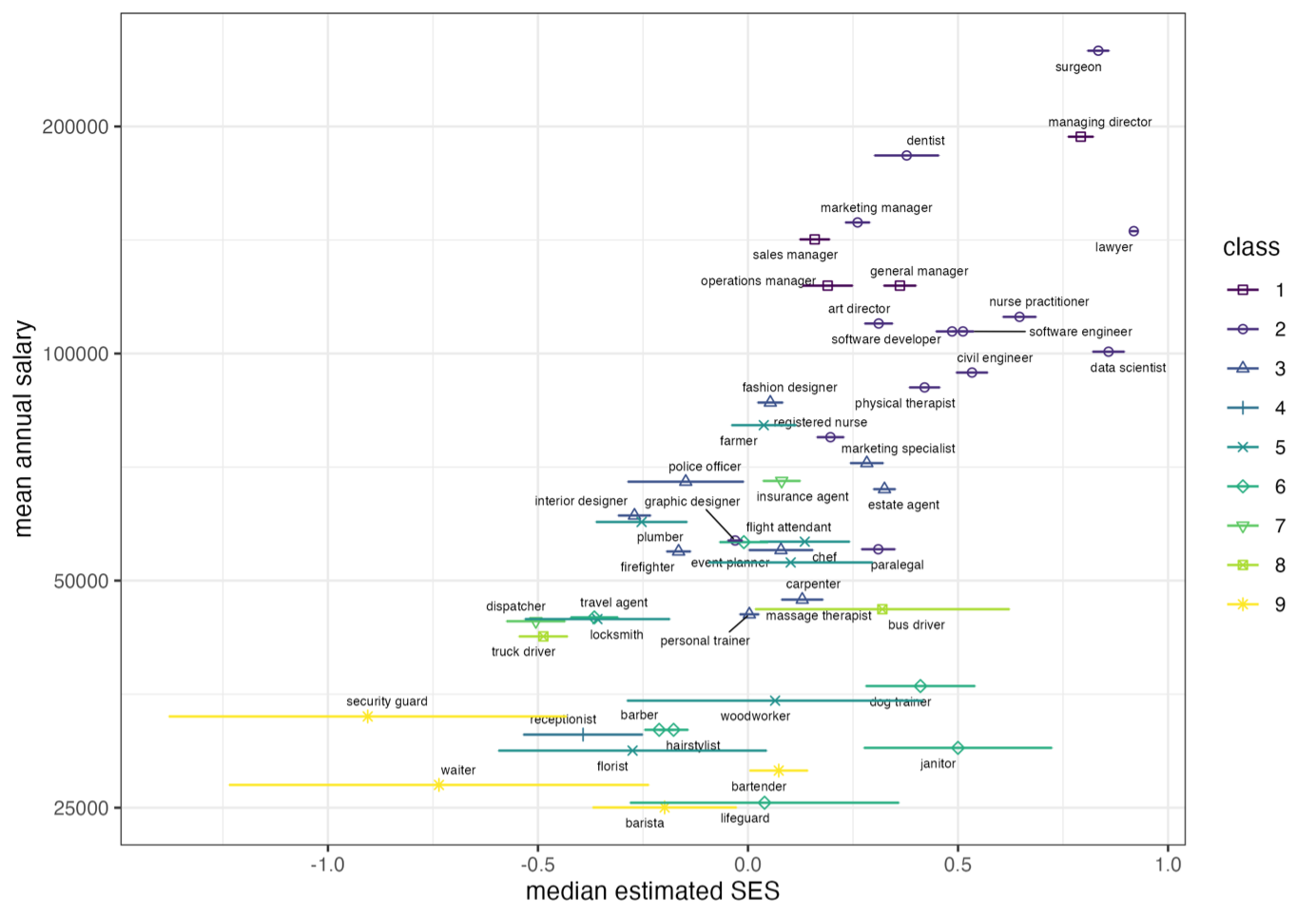}
	\caption{Relation between median estimated SES, mean salary, and occupational class for a set of 50 common job titles, estimated over 42,099 Twitter users who mention one of the titles in their profile description.}
	\label{fig:fig4}
\end{figure}

Figure \ref{fig:fig4} depicts the association between the median estimated SES of users for each job title and the job title’s mean annual salary and occupational class. The salary is logarithm scaled with base 10, the bars show standard errors for the median estimated SES, and the colours represent the occupational social class, where higher number means lower class. There is a clear positive trend, where jobs with a higher median estimated SES tend to have a higher mean annual wage. Jobs with the same class also tend to cluster. From bottom left to top right, there is a discernible trend from low salary, class, and estimated SES to higher salary, class, and estimated SES. Statistical tests show that the Spearman’s rank correlation between the median estimated SES and mean annual salary is $0.673$ ($p < 0.001$). The Spearman’s rank correlation between the median estimated SES and occupational class is $-0.640$ ($p < 0.001$). As a reference, in our sample, the Spearman’s correlation between mean annual salary and class is $-0.840$ ($p < 0.001$). Although the correlations between our estimated SES and salary or class are not as high as the well-established correlation between salary and class, they are sufficiently strong to validate the proposed method at the aggregate level.

Nevertheless, as the error bars in Figure 4 show, there are large variations of estimated SES for some job titles, especially at the lower end of SES. Individual earnings for the same job title vary depending on US State, urban setting, business size, etc. At the aggregate level, the effects of these variations may cancel out by the large number of users selected for each title, but the effects will be more palpable at the individual level. Therefore, we next use individual-level SES data to further validate our estimates.

\subsection{Validation of user SES with survey data}
To test the method with better ground truth data, we identify a small sample of the Twitter brand followers who report their income and educational level in a survey. The survey data were provided by Guess et al. (2021), who recruited 1,551 respondents from the YouGov U.S. Pulse panel, 471 of whom shared their Twitter data. Restricting the sample to users who follow at least one of the brands from our sample, we were left with 200 users whose SES we can estimate with our method. For these 200 users, we have their self-reported highest educational level as ordinal variable from one to six, coded as 1: No high school, 2: High school graduate, 3: Some college, 4: Two-year college, 5 Four-year college, and 6: Post-graduate. For 182 users, we also have their income level data coded as an ordinal variable from one to 16, starting from Less than \$10,000, then going in increments of \$10,000 up to \$80,000, after which the categories start from \$100,000, \$120,000, \$150,000, \$200,000, \$250,000, \$350,000, and finally, \$500,000 or more. Using these data, visually presented in Figure 5, we estimate the Spearman correlation between estimated SES and educational level to be $0.269$ ($p < 0.001$) and the one between estimated SES and income level to be $0.188$ ($p < 0.05$). As a reference, the Spearman correlation between income and education in the sample is $0.455$ ($p < 0.001$), which is surprisingly low. If we restrict the survey sample to Twitter users who follow at least two or three accounts, the correlations with education improve ($0.259$, $p < 0.001$, $N = 147$ in the case of two accounts; $0.344$, $p < 0.001$, $N = 111$ for three accounts) but weaken for income ($0.137$, $p = 0.117$, $N = 131$ for two accounts; $0.156$, $p = 0.117$, $N = 102$ for three accounts). These results suggest that our method successfully captures information relating to SES and specifically, captures education better than income. Figure 5 reveals that the model is particularly successful in identifying highly educated individuals with high income. Nevertheless, there appears to be a significant amount of noise or, possibly, unrelated demographic information. Ideally, we would have access to larger survey data to identify for whom the method underperforms. At the very least, we should establish that the proposed method captures SES constructs better than other associated demographic variables. This is what we do next. 

\begin{figure}[ht]
	\centering
	\includegraphics{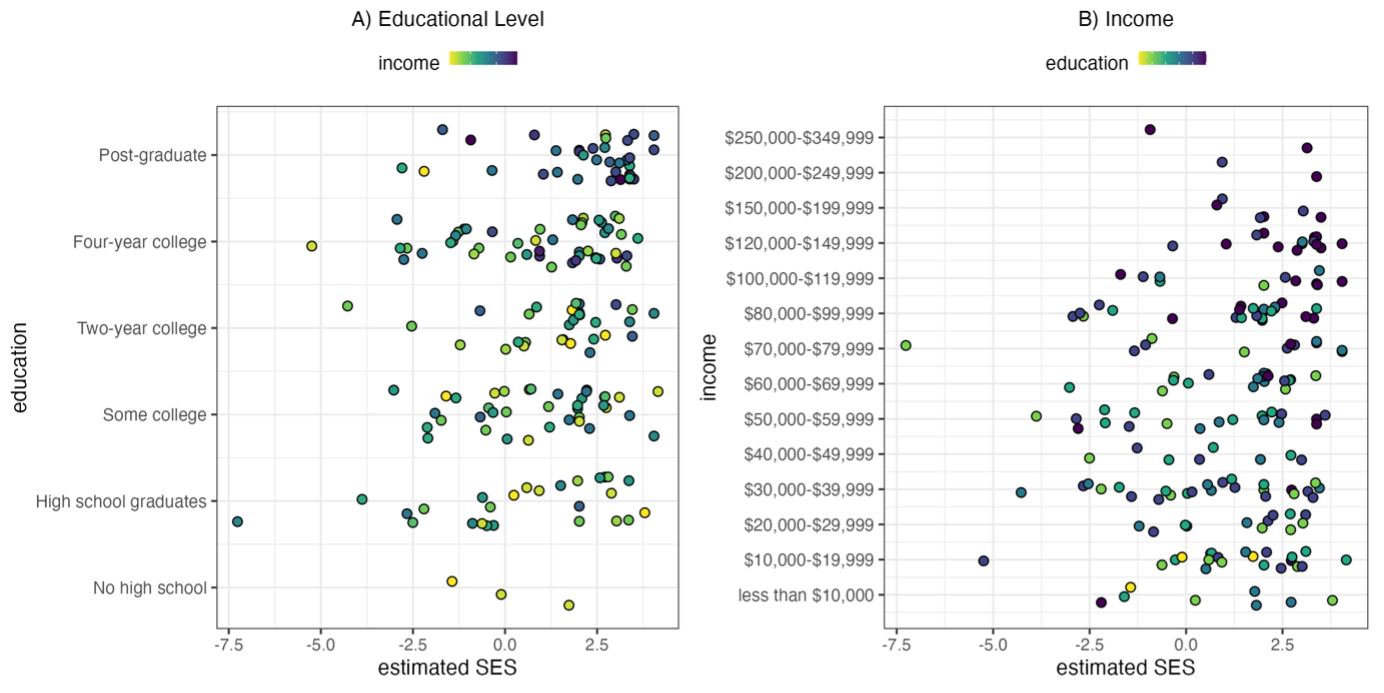}
	\caption{Relation between estimated SES and A) educational level and B) income for 200 (182 for B) survey respondents who follow at least one of the 339 brands on Twitter. The y-axis values are plotted with noise to improve visibility.}
	\label{fig:fig5}
\end{figure}

\subsection{Divergent validity}
So far, we focused on convergent validity, utilising multiple sources of data to establish that the estimates are correlated with other proxies for the theoretical concept of SES. To further establish the validity of the measurement method, we also provide evidence for divergent validity, demonstrating that the estimated SES does not capture other demographic variables related to SES better. 

First, with similar data from the Facebook Marketing API, we analyse the associations between the estimated SES and the proportion of urban users, male/female users, and users in different age groups. \footnote{The divergent validity analysis was conducted two years after the convergent validity analysis, during which period the Facebook Marketing API changed the searchable terms and some brands went bankrupt. Therefore, the number of brands with suitable audience data dropped from 328 to 295. The details of the unavailable brands are included in Supplementary Table 4. Additionally, the API now does not return one number for the estimated target audience size but returns lower bound and upper bound. Here, we present the results using the average between lower and upper bound. The results from the average, lower, and upper bound are essentially the same.} The estimated SES of the brands is very weakly associated with the proportion of urban users ($\rho = 0.114$, $p = 0.050$) and not associated with the proportion of male ($\rho = 0.034$, $p = 0.558$) nor female ($\rho = -0.037$, $p = 0.532$) users. These results suggest our SES measure for the brands is not capturing urban/rural nor gender disparity. The estimated SES of the brands has significant but weak positive associations with the proportion of users in younger age groups: Spearman correlation coefficients of $0.172$ ($p < 0.01$) for age 18-24, $0.199$ ($p < 0.001$) for 25-34, and $0.136$ ($p < 0.05$) for 35-44. Conversely, the estimates have weak negative associations with older age groups: Spearman correlation coefficients of $-0.135$ ($p < 0.05$) for age 45-54, $-0.224$ ($p < 0.001$) for 55-64, and $-0.117$ ($p < 0.05$) for 65 and above. Although statistically significant, the associations between estimated SES and age are much weaker than education and hence, we can conclude that the estimated SES for the brands captures education better than age. 

Second, we test the correlation between estimated SES and political ideology, as measured by Barberá et al.'s \citeyear{barbera_tweeting_2015} method. For the 150,011 informative users whose Twitter followings are still available in November 2022, the Spearman correlation between estimated SES and political ideology is $-0.114$ ($p < 0.001$), where positive values for ideology mean conservative-leaning. The large sample size contributes to the statistical significance, but the correlation is weak. Thus, although we use the same method and several overlapping official Twitter accounts (mainly news), the modification we propose no longer reflects political ideology at the individual level.

Third, using again data from \citet{guess_consequences_2021}, we test the associations between estimated SES and related demographic variables available in the survey: age, gender, political ideology, and race. The estimated SES is not significantly associated with any of the variables tested. For the 195 participants with available demographic data, the Spearman correlation with age is $0.106$ ($p = 0.139$) and the t-test between male and female is $0.741 $($p = 0.460$). Similarly, there is no significant difference in estimated SES between the four racial groups (White, Black, Hispanic, Asian/other) categorised in the survey, regardless of whether we use an analysis of variance test ($p = 0.871$) or pair-wise t-tests. For the 189 participants with self-reported political ideology (a scale from 1 to 5), the correlation between estimated SES and political ideology is $-0.079$ ($p = 0.279$). As a reference, in the sample, education and income are also not significantly associated with any of the four variables (detailed results are available in Supplementary Table 5). Further, regression analyses, presented in Supplementary Table 6, show that controlling for age, gender, political ideology and race, there are still significant correlations between estimated SES and education ($p < 0.001$) and between estimated SES and income ($p < 0.05$).

Overall, the estimated SES has insignificant or weak associations with related demographic variables such as age, gender, race, political ideology, and urban/rural residence, while the correlations between the estimated SES and established SES proxies, including education, income, and occupational class, are significant and much stronger. Combined together, the results of the analyses of convergent and divergent validity provide a strong case for the validity of the proposed method. 

\section{Discussion}
This study presents a method for estimating Twitter users’ SES from the consumer and media brands they follow. The method is adapted from a widely used approach to measuring Twitter users’ political ideology. Compared to previous attempts to estimate SES from social media data, the proposed method is built on behavioural assumptions that can be linked to classical sociological theory, requires only a basic understanding of a common dimensionality reduction technique, and provides estimates for millions of individuals while only using minimal, easily available and obtainable data, open-source off-the-shelf software programs, and modest computational power. We applied the method using 339 popular US brands to estimate the SES of almost 3.5 million Twitter users. We then brought in additional data, including advertisement audience statistics from Facebook, user profile information from Twitter, and survey sample responses, to validate the accuracy of the estimates with the standard SES proxies of education, occupational class, and income and confirm their dissociation from other demographic variables known to be related with SES.

The results suggest that the proposed measure of SES for Twitter users is promising. The measure works well at the aggregate level but needs finetuning with better validation data for more precise individual estimates. The estimated SES for the brands correlates reasonably well with the educational level of their audience and aligns intuitively with general brand perceptions. Aggerated for a selected group of job titles, the estimated SES for users is also strongly associated with annual mean salary and occupational class. At the individual level, the SES estimates are significantly associated with education and income, but the correlations are relatively weak. Further, for both brands and individuals, the SES estimates are not, or at best much weakly, associated with related demographic variables, including age, gender, race, urban/rural residence and political ideology. Overall, the significant associations between the estimated SES and the traditional SES indicators and the insignificant or weak associations with other demographic variables at both the aggregate and individual levels support the underlying principle of the proposed method and justify further efforts to refine it at the individual level.

Nevertheless, we interpret the results with some further reflections on the theoretical assumptions and methodological choices we made. The main principle of the proposed method is that Twitter users manifest their economic and cultural interests with the brands they follow and hence these brands can inform us about their SES. We note that following a brand on Twitter does not involve any economic costs and does not necessarily imply real material consumption. Yet, no economic cost does not mean no cost at all. Users have finite ability to process information and divide attention on Twitter \citep{hodas_how_2012}. Following an account populates one’s newsfeed with updates, displacing other relevant information and this is particularly the case for official accounts managed by professionals who regularly produce content. In other words, while clicking to follow Whole Food’s Twitter account is just as effortless as clicking to follow Aldi’s account, there are direct and opportunity information costs associated with remaining a follower.       

Unconstrained by cost, Twitter users may follow brands for many possible reasons that are not relevant to economic or cultural interests, e.g., out of curiosity or by mistake. We certainly cannot assume that all brand followings are based on economic and cultural interests associated with SES, but we propose that the dominant trend is related to SES. The validation results indeed indicate that SES has a significant role to play. This observation also aligns with evidence that the digital world reflects and even reproduces the existing cultural boundaries of the physical world regarding people’s interests in restaurants, music, films, museums, and galleries \citep{airoldi_techno-social_2021, goldberg_what_2016, mihelj_culture_2019}, and even more so, politics \citep{bail_exposure_2018, tucker_social_2018}. The basic principle behind the proposed method is to exploit these digital cultural and lifestyle boundaries to obtain information about individuals, which can then be used in research that challenges them. 

Another related objection is that following a brand on Twitter might be aspirational and reflect desired, rather than actual SES. We know that, on the one hand, people universally desire higher social status \citep{anderson_is_2015, fiske_envy_2011} and on the other, online users strategically orchestrate online personas and actively manage their self-presentation online \citep{schlenker_strategic_2000}. However, since followed accounts are not easily and directly observable on a user’s profile, they are unlikely to be employed solely as status signals. A user can signal status with the accounts they follow only if they actively retweet or @-mention them, so future work could analyze such activity to estimate the extent to which followings are status-seeking rather than status-reflecting. Additionally, we note that the unsupervised learning method we employ is agnostic to \textit{a priori} brand associations or expectations. The method positions the brands according to their co-followings and it can thus place an expensive brand towards the low-SES end of the spectrum if its audience on Twitter tends to consist of consumer-hopefuls rather than actual consumers. Nevertheless, we recognize that strategic self-presentation may be idiosyncratic and as such, it will inevitably introduce noise to the individual estimates.  

Finally, we note that the weak signal at the individual level the method achieves should be interpreted in light of the natural limits of predictability of human behavior social scientists face \citep{hofman_prediction_2017, song_limits_2010}. As we discussed above, besides actual SES, strategic self-presentation, unknown personal motivations, other demographic characteristics, peer effects, and situational factors could dictate whether a specific individual follows a brand. This inevitable degree of idiosyncrasy and complexity means that the salient effect of SES may only manifest at the aggregate level, but dissolve at the individual level. A recent large-scale mass collaboration scientific project shows that, even with high quality data and sophisticated methods, the predictability of individual life outcomes is still very low \citep{salganik_measuring_2020}. We soberly recognize that similar natural limits likely constrain the measurement of individual SES of Twitter users from their expressed cultural interests and consumer preferences. 

Despite these inherent limitations, we see a huge potential in further efforts to validate, refine, and apply the proposed method. The next natural step is to link richer survey data of a larger sample with Twitter user profiles. This step involves extra resources and additional methodological and ethical issues \citep{baghal_linking_2021, stier_integrating_2019} but the resulting linked data could contribute in multiple ways. First, the data will allow re-validating the proposed method, disentangling demographic factors that strongly influence the SES estimates, and quantifying the extent to which the measures correspond to actual versus desired SES. Second, the data can be used to fit supervised learning models for estimating SES to improve the proposed unsupervised method but also compare the strengths and weakness of different methods, examine the inherent limits to the predictability of individual SES, and recommend suitable methods for different situations. 

One way in which a supervised learning model on a linked survey data could help improve the proposed method is by refining the consumer domains and official accounts to include in the estimation. The included official accounts determine whether correspondence analysis indeed captures the variations in SES. In this study, we consulted a variety of sources to select a group of brands that represent a wide range of economic and cultural interests, but this selection could be improved with a more data-driven approach. Although there are numerous studies on the link between taste and social status, especially following Bourdieu’s \citeyearpar{bourdieu_distinction_1984} work \citep[e.g.,][]{alderson_social_2007, chan_social_2007, gerhards_social_2013, katz-gerro_cultural_1999, peterson_understanding_1992, reeves_how_2019}, there is limited research on the specific brand preferences of people in different SES. The brands themselves rarely disclose their audience demographics. Future research would benefit from a comprehensive analysis of the relation between SES and specific interests using sources such as the Facebook Marketing API and other mobile or web tracking data, linking it to previous research on SES and taste. Such research will provide not only a more informed selection of the official accounts to include in the model but also a more comprehensive picture on SES, taste, and habitus.

Although we carefully considered the six domains we chose (supermarkets and department stores, clothing and speciality retailers, chain restaurants, newspapers and news channels, sports, and TV shows), this set is not necessarily comprehensive. One may argue that news sources, sports, and TV shows are very reductive parts of cultural interests that people express on Twitter, and that artists, musicians and influencers should also be included. Indeed, the current set of domains carries the danger of reducing cultural capital to consumerism, especially with its focus on “brands”. For this initial attempt, we took a more conservative approach and chose consumer brands that combine economic and cultural interests, avoiding accounts related to art and music. Music and art form the core of cultural capital, but also fuel intense debates about the link between cultural capital and SES. The highbrow-vs-omnivore debate around cultural capital, where art and music activities are often used as empirical evidence, is ongoing and active \citep{chan_understanding_2019, goldberg_mapping_2011, de_vries_what_2021}. We thus expect the contribution of musicians and artists to SES estimation in our method would be less informative and less interpretable. Nevertheless, this constitutes an empirically testable hypothesis that future work could explore. Work that adds artists, musicians and other related accounts to the proposed model could potentially both benefit our method and contribute to the ongoing highbrow-vs-omnivore debate. 

Despite the mentioned limitations and aspects for improvement, the proposed method carries an enormous promise for social science research. The method provides SES estimates on a continuous scale that are operationally easy to use and theoretically interpretable. Social scientists could combine these SES estimates with digital trace data on behaviours, communication patterns, and social interactions to study inequality, health, and political engagement, among many other topics. For instance, one can link our measure of SES, which captures cultural and economic capital, to indicators of social capital inferred from social relations and interactions on Twitter and explore how the different forms of capital combine to contribute to socioeconomic inequality. Specifically, we can now study the effects of social networks on inequality, as discussed by \citet{dimaggio_network_2012}, with new data, in a different context, and on a significantly larger scale.

The SES estimation method we propose here opens myriad new avenues for academic research on Twitter and similar social network platforms. We used Twitter due to its popularity and convenient API, but the principle of our method can be applied to many other online platforms. For example, future research can use the interests expressed by following or liking certain topics or accounts to estimate the SES of users on platforms such as Reddit and Quora and then link SES to behaviours, opinions, and knowledge expressed on those platforms.

\large{\textbf{Acknowledgements}}

The authors are grateful to Andrew Guess, Pablo Barberá, Simon Munzert, and JungHwan Yang for sharing data, to Eleanor Power and Blake Miller for valuable detailed feedback, and to the reviewers for their constructive comments. 

\large{\textbf{Authors’ Note}}

Data, codes, and a README file detailing what is included and how to reproduce the published results are available on Figshare (\url{https://doi.org/10.6084/m9.figshare.22007000.v1}) and GitHub (\url{https://github.com/yuanmohe/Twitter_SES}).

\bibliographystyle{plainnat}
\bibliography{references}

\end{document}